\documentclass[runningheads]{llncs}
\usepackage{graphicx}
\begin{document}
%
\title{Improved Topic modeling in Twitter through Community Pooling}
%
%
\author{Federico Albanese\inst{1,2}\orcidID{0000-0001-7140-2910} \and
Esteban Feuerstein\inst{1,3}\orcidID{0000-0003-2985-810X}}
\authorrunning{Albanese et al.}
%
\institute{Instituto en Ciencias de la Computaci\'on, CONICET - Universidad de Buenos Aires, Argentina \and
Instituto de C\'alculo, CONICET- Universidad de Buenos Aires, Argentina\\ \and
Departamento de Computaci\'on, Universidad de Buenos Aires, Argentina\\
\email{falbanese@dc.uba.ar}}
\maketitle              
\begin{abstract}
Social networks play a fundamental role in propagation of information and news. Characterizing the content of the messages becomes vital for different tasks, like breaking news detection, personalized message recommendation, fake users detection, information flow characterization and others. However, Twitter posts are short and often less coherent than other text documents, which makes it challenging to apply text mining algorithms to these datasets efficiently. Tweet-pooling (aggregating tweets into longer documents) has been shown to improve automatic topic decomposition, but the performance achieved in this task varies depending on the pooling method.

In this paper, we propose a new pooling scheme for topic modelling in Twitter, which groups tweets whose authors belong to the same \emph{community} (group of users who mainly interact with each other but not with other groups) on a user interaction graph. We present a complete evaluation of this methodology, state of the art schemes and previous pooling models in terms of the cluster quality, document retrieval tasks performance and supervised machine learning classification score. Results show that our Community polling method outperformed other methods on the majority of metrics in two heterogeneous datasets, while also reducing the running time. This is useful when dealing with big amounts of noisy and short user-generated social media texts. Overall, our findings contribute to an improved methodology for identifying the latent topics in a Twitter dataset, without the need of modifying the basic machinery of a topic decomposition model. 

\keywords{Topic modelling  \and Community detection \and Twitter \and Text mining \and Text clustering}
\end{abstract}

\section{Introduction}
Characterizing texts based on their content is an important task in machine learning and natural language processing. Latent Dirichlet Allocation (LDA) is a generative model for unsupervised topic decomposition \cite{blei2003latent}. Documents are represented as random mixtures over topics with a Dirichlet distribution, and each topic is characterized by a distribution over words. LDA has been widely used for topic modeling in different areas such us medical science \cite{paul2011you}, political science \cite{cohen2013classifying}, social computer science \cite{pinto2019quantifying} and software engineering \cite{gethers2010using}.

In practice, content analysis on microblogging services can be particularly challenging due to short and often vaguely coherent text \cite{mehrotra2013improving,ma2019time}. Given the fact that Twitter has become a platform where a tremendous amount of content is generated, shared and consumed, this problem become of interest for the scientific community.
Hong presented an intuitive solution to this problem: tweet pooling (making longer document by aggregating multiple tweets) \cite{hong2010empirical}. Tweet-pooling has been shown to improve topic decomposition, but the performance varies depending on the pooling method \cite{hong2010empirical,mehrotra2013improving,ma2019time,alvarez2016topic,ollagnier2019network}. For example, Mehrotra et al. \cite{mehrotra2013improving} extended this idea by pooling all tweets that mention a given hashtag. More pooling techniques are described in detail in Section 2.

In this paper, we propose a novel pooling techniques based on community detection on graphs. Previous works stated that LDA has problems with sparse word co-occurrence matrix \cite{ma2019time} and showed that users in a community tweet mostly about two or three particular topics \cite{albanese2020predicting}. Based on these issues, we propose a community pooling method which groups tweets whose authors belong to the same community on the retweet network, increasing the length of each document and reducing the total number of documents.
We compare the schemes in terms of clustering quality, document retrieval, machine learning classification tasks and running time and we empirically show that this new scheme improves the performance over previous methods in two heterogeneous Twitter datasets.

The remainder of this work is organized as follows: In Section 2 we describe the different pooling schemes for topic models and propose a novel method. In Section 3 we describe the datasets that we used to test our method. In section 4, we define the experiments and evaluation metrics that we use to measure the performance of all pooling schemes. In section 5, we show the results of the experiments. Finally, we interpret the results in the Conclusions section.

\section{Tweet pooling for topic models}

Microblog messages are very short texts. In particular Twitter posts are only $280$ characters or shorter. Consequently, using each tweet as an individual document does not present adequate term co-occurence data within documents \cite{mehrotra2013improving}. This induced the idea that aggregating similar tweets gives place to larger documents and better LDA topic decomposition.  
In this section, we present a new pooling method for topic modeling based on community detection and describe five other methods proposed in the literature which were used for comparison.

\textbf{Tweet-pooling (Unpooled):} The default approach which treats each tweet as a single document. 

\textbf{Author-Pooling:} All tweets authored by a single user are aggregated in a single document. The number of documents is equal to the number of users. This pooling method outperforms the Unpooled scheme \cite{hong2010empirical}.

\textbf{Hashtag pooling:} In this scheme, a document consists of all tweets that mention a given hashtag. A tweet that contains multiple hashtags appears in several documents. Tweets without hashtags are considered as individual documents. It has been shown that aggregating tweets this way outperforms the baseline scheme and user-pooling in some metrics and for some datasets \cite{mehrotra2013improving}.

\textbf{Conversation pooling:} A document consists of all tweets in a conversation tree (i.e. a tweet, all the tweets written in reply to it, the replies to the replies, and so on). This schemes aggregate tweets form different authors and with multiple hashtags that belong to one conversation \cite{alvarez2016topic}.

\textbf{Network-based pooling:} Twitter users are grouped together if they reply or are mentioned in a tweet or in replies to a tweet. Each single document consists of all tweets of a group of users. In contrast to Conversation pooling, only direct replies to an original tweet are considered since a conversation can shift its topic in time. This pooling scheme showed better results than the previous methods in most (not all) tasks and datasets \cite{ollagnier2019network}.


\textbf{Community pooling:} In this novel scheme, a retweet graph is defined in terms of $G = (N,E)$, where users are the nodes $N$, and retweets between them are edges $E$ \cite{aruguete2018time}. Since a user can retweet multiple times other user's tweets, the edges are weighted. A community in a social network is a group of users who mainly interact with each other but not with other groups. We determine these communities using the Louvain method for community detection \cite{blondel2008fast}, which seeks to maximize modularity by using a greedy optimization algorithm. Therefore, each community clusters users by their interactions.
In our novel pooling method, we group in one document all the tweets authored by all users in each community. Therefore, there are as many documents as communities in the retweet network. Compared with the majority of the previous schemes, the number of words in a document is bigger and the number of documents is smaller, resulting in a denser word co-occurrence matrix, which is beneficial to LDA algorithm \cite{alvarez2016topic}.

\section{Twitter dataset construction}

In order to evaluate the schemes in different scenarios and show the robustness of the methodology, we used two diverse datasets. Our experiments used data from Twitter Streaming API \footnote{https://developer.twitter.com/en}. Similarly to previous works, we constructed two diverse datasets collecting tweets containing different queries and each tweet was labeled by the query that retrieved it \cite{ollagnier2019network,hong2010empirical,mehrotra2013improving,al2019events}. We removed all tweets that were retrieved by more than one query, so as to preserve uniqueness of the tweet labels, which was important for our analysis. Besides, we prepossessed the tweets by lower-casing and removing stop-words. All tweets are in English. The two datasets are:

\textbf{Generic Dataset:} 115,359 tweets from December $15^{th}$ to December $16^{th}$, 2020, concerning a wide range of themes and collected using the following queries (percentage of
tweets retrieved by each query): music ($36.78\%$), family ($23.94\%$), health ($17.21\%$), business ($14.90\%$), movies ($4.70\%$), sports ($2.44\%$).

\textbf{Event Dataset:} 328,452 tweets from January $20^{th}$, 2021. A dataset composed of tweets belonging to a particular event: US president Biden inauguration day. We used the following queries: Biden ($69.45\%$), joebiden ($21.75\%$), kamalaharris ($4.74\%$), inauguration2021 ($4.04\%$).

\section{Evaluation}

As there is no standard way for evaluating topic models, previous works evaluated the proposed pooling methods using different metrics or tasks. In order to present a complete and exhaustive analysis, in this work we evaluate the schemes by the multiple metrics used in the different previous works: topic clustering metrics (Purity and Normalized Mutual Information) \cite{alvarez2016topic,hajjem2017combining,mehrotra2013improving,quezada2019lightweight,ollagnier2019network,akhtar2019user}, a supervised machine learning classification task \cite{giorgi2018remarkable,hong2010empirical},
a document retrieval task \cite{alvarez2016topic,al2019events} and overall running time \cite{alvarez2016topic}. We briefly explain each of them.

\textbf{Purity:} We define each cluster as a topic and assign the tweets to their corresponding mixture topic of highest probability (a quantity estimated with LDA). The purity of a cluster measures the fraction of tweets in a cluster having the assigned cluster query label \cite{schutze2008introduction}. 
Formally, let $T_i$ be the set of tweets in LDA topic cluster $i$ and $Q_j$ be the set of tweets with query label $j$. Let $T = \{T_1, T_2, ..., T_{|T|}\}$ be the set of size $|T|$ of all $T_i$ and let $Q = \{Q_1, Q_2, ..., Q_{|Q|}\}$ be the set of size $|Q|$ of all $Q_j$. 
Then, the purity is defined as follows:

\begin{equation}
\textrm{Purity}(T, Q) =  \frac{1}{|T|} \sum_{i \in \{1...|T|\}} max_{j \in \{1...|Q|\}} |T_i \cap Q_j|
\end{equation}

A higher purity score reflects a better cluster representation and a better LDA decomposition.

\textbf{Normalized Mutual Information (NMI):} NMI measures the cluster quality using information theory and it is formally defined as follows:

\begin{equation}
\textrm{NMI}(T, Q) =  \frac{2I(T,Q)}{H(T)+H(Q)} 
\end{equation}

where $I(\cdot , \cdot)$ is the mutual information and $H(\cdot)$ is the entropy, as defined in \cite{schutze2008introduction}. NMI's minimum and maximum values are resp. 0 when labels and clusters are independent sets and 1 when cluster results exactly match all labels.

\textbf{Supervised machine learning classifying task:} For the supervised machine learning task, we follow a basic machine learning classifying evaluation scheme \cite{hong2010empirical}. We separate the dataset in two (train and test), train a classifier with the first one and evaluate on the second one. The first $80\%$ of tweets (according to the time their were posted) were assigned to the train set and the other $20\%$ to the test set. 
For this task, we train a naive Bayes classifier \cite{muller2016introduction} and reported F-Measure (F1 score) on the test set.

\textbf{Document retrieval task:} We also evaluate the topic decomposition of the different pooling methods on a document retrieval task, using the same train-test split as the supervised classifier task. We use each tweet in the test set as a query and return the most similar tweets from the train set, according to their LDA topic decomposition. If the retrieved tweet has the same query label, we consider it relevant. More concretely, the methodology is as follows: we apply LDA using the different pooling techniques on the train set, for each tweet in the test set calculate its topic decomposition, compute the cosine similarity between its topic decomposition and the topic decomposition of all tweets in the train set and retrieve the top 10 most similar train tweets. Then, we calculate the F1 score in order to know if the the categories of the retrieved tweets match the category of the test tweet. This task recreates a scenario of recommending content based on previous tweets.

\textbf{Running time:} The measured time (in seconds) includes tweet pooling (aggregating the tweets in different documents) and the LDA topic modeling, which varies depending on the total number of documents of each pooling methods. 

All experiments were run using the same hardware on a GTX 1080 NVIDIA graphic card.

\section{Results}

In this section we show and discuss the results of our evaluation. For each pooling scheme, we replicated the training workflow used in the literature and used an LDA model with 10 topics \cite{mehrotra2013improving,ollagnier2019network}. 
As we mentioned earlier, previous works showed that having denser co-occurence matrix (fewer documents with more words each) is beneficial to LDA \cite{alvarez2016topic}. Table \ref{table_corpus} reports the corpus characteristics and shows how our proposed model drastically reduced the number of documents and increased the number of words per document.

\begin{table}
\caption{Document characteristics for different pooling schemes and datasets.}\label{table_corpus}
\begin{tabular}{|l|l|l|l|l|l|l|}
\hline
Scheme           & \multicolumn{2}{l|}{\# of docs}         & \multicolumn{2}{l|}{Max \# of words/doc}  & \multicolumn{2}{l|}{Mean \# of words/doc}  \\
           & generic  & event      & generic & event & generic & event   \\
\hline
Unpooled         &   115,359  & 328,452  &  783  &  1,023 &  137 &  128   \\
Author           &   36,526  & 87,883  &  36,029  &  11,240 &  369 &  273 \\
Hashtag          &   34,624  & 59,388  &  820,689  &  3,736,132 &  8295 &  173952  \\
Conversation     &   35,484  & 67,276  &  12,480  &  41,024 &  141 &  130  \\
Network-based    &   36,882  & 88,314  &  59,195  &  90,391 &  385 &  277 \\
Community        &   24,657  &  31,303  & 2,077,085 & 5,284,617 & 874 & 1379 \\
\hline
\end{tabular}
\end{table}

The results of the experiments can be seen in table \ref{table_results1}. The best performances are marked in bold. The table shows that Community pooling has the best performance of all examined methods in all metrics for the Generic Dataset, and in all metrics except the retrieval task for the Event Dataset.

\begin{table}
\caption{Results for different pooling schemes and datasets.}\label{table_results1}
\begin{tabular}{|l|l|l|l|l|l|l|l|l|l|l|l|l|}
\hline
Scheme           & \multicolumn{2}{l|}{Purity}         & \multicolumn{2}{l|}{NMI}  & \multicolumn{2}{l|}{Classification}  & \multicolumn{2}{l|}{Retrieval}     & \multicolumn{2}{l|}{Running time}  \\
           & generic  & event      & generic & event & generic & event  & generic & event & generic & event\\
\hline
Unpooled         & 0.664          & 0.733 & 0.436  & 0.110  &  0.814  &  0.843 &  0.837 &  0.893  & 137  &  388\\
Author           & 0.696          & 0.736 & 0.374  & 0.149  &  0.798  &  0.859 &  0.839 &  0.900 &  429  &  926\\
Hashtag          & 0.724          & 0.719 & 0.383  & 0.066  &  0.779  &  0.762 &  0.839 &  0.869  &  1,737& 17,758\\
Conversation     & 0.658          & 0.733 & 0.436  & 0.110  &  0.814  &  0.843 &  0.835 &  0.908  & 738  & 1,569\\
Network-based    & 0.695          & 0.736 & 0.372  & 0.149  &  0.798  &  0.859 &  0.840 &  \textbf{0.910} &  1131 & 2,841\\
Community        & \textbf{0.780} & \textbf{0.779} & \textbf{0.439}  &  \textbf{0.310}  & \textbf{0.827} & \textbf{0.889} & \textbf{0.843} & 0.868 & 141  & 340\\
\hline
\end{tabular}
\end{table}

Our methodology obtained the best cluster quality, having the highest Purity and NMI scores. Also our experiments showed that Community pooling outperformed the previous schemes in the supervised classification task,  indicating that this topic decomposition was a good descriptor of the query label. Regarding the document retrieval task, this evaluation considers small changes in the topic decomposition of a tweet, since it uses the cosine similarity between this decomposition instead of only taking into account the most likely topic as we did before with the clustering metrics. The results indicate that Community pooling had the best performance in a generic dataset where the topics of the labels (``family'', ``health'' or ``business'') differentiate from each other. In contrast, we found that the Network-based method has a better score in this task for the event dataset, where the labels are closely related (``joebiden'' and ``kamalaharris''). Community pooling has better performance on all tasks and datasets, with the only exception of the retrieval task on the event dataset.

Finally, Community pooling had the best time performance among all pooling methods. From the fact that LDA time complexity depends on the number of documents \cite{scikit-learn} and Community pooling considerably reduced the number of documents by pooling together in a single document all the tweets posted by users of each community (see table \ref{table_corpus}), it follows that our proposed method was faster than all other aggregation techniques (less than half the running time). 

\section{Conclusions}

We presented a new way of pooling tweets in order to improve the quality of LDA topic modeling on Twitter, without requiring any modification of the underlying LDA algorithm. The proposed Community pooling uses the users' interaction information and aggregates into a single document all tweets of the users that belong to a community in the retweet network.

Our method was evaluated and compared with multiple pooling techniques on different task including clustering quality, a supervised classification problem and a retrieval tasks. The results on two heterogeneous datasets indicate that the novel Community based pooling outperforms all other pooling strategies in all tasks and metrics, with the only exception of the retrieval task on the event dataset. Also, the running time analysis shows that Community pooling has a significant improvement in time performance in comparison with previous pooling methods, due to its capacity of reducing the total number of documents.
Future work includes further testing with other datasets from different social media.

%
%
%

\end{document}